\documentclass[smallextended]{svjour3}  

\usepackage[authoryear]{natbib}

\usepackage{graphicx} 
\usepackage{amsmath}
\usepackage{enumerate}
\usepackage{threeparttable}
\usepackage{booktabs} 
\usepackage{calc}
\usepackage{pifont}
\usepackage{color}
\definecolor{light-gray}{gray}{0.80}
\definecolor{grannysmithapple}{rgb}{0.66, 0.89, 0.63}
\definecolor{green(html/cssgreen)}{rgb}{0.0, 0.5, 0.0}
\definecolor{brightmaroon}{rgb}{0.76, 0.13, 0.28}

\usepackage{framed}
\usepackage{wrapfig}
\usepackage{graphicx}
\usepackage{fancyvrb}
\usepackage[dvipsnames]{xcolor}
\usepackage[most]{tcolorbox}
\usepackage{natbib}
\usepackage{amsmath,amssymb,amsfonts}
\usepackage{textcomp}
\usepackage{booktabs}

\usepackage{tikz}
\usetikzlibrary{shapes,arrows}
\usetikzlibrary{intersections}
\usetikzlibrary{positioning}
\usetikzlibrary{automata, positioning, arrows}

\usepackage{url} 
\usepackage{xspace}
\usepackage[normalem]{ulem}
\usepackage{pifont} 
\usepackage{tabularx}
\usepackage{subcaption}
\usepackage{listings}
\usepackage{csvsimple}

\newcommand\encircle[1]{%
  \tikz[baseline=(X.base)] 
    \node (X) [draw, shape=circle, inner sep=-1, fill=black, text=white, 
    minimum size=.4cm
    ] {\strut #1};%
}

\usepackage[utf8]{inputenc}
\usepackage{algpseudocode}
\usepackage{helvet}
\usepackage{algorithm}

\usepackage{hyperref}
\usepackage{graphicx}

\usepackage{subcaption}
\usepackage{enumitem} 
\usepackage{booktabs}
\usepackage{multirow}
\usepackage{makecell}
\usepackage{pgfplots}
\usepackage[utf8]{inputenc}

\usepackage{verbatim}

\renewcommand{\theenumi}{\Alph{enumi}}

\newenvironment{sansserifalgoritm}{%
    \fontfamily{\sfdefault}\selectfont
    \begin{algorithm}
}{%
    \end{algorithm}
}

\definecolor{bluetext}{RGB}{0,0,255}
\algrenewcommand\algorithmicfunction{\textcolor{bluetext}{\textbf{function}}}
\algrenewcommand\algorithmicwhile{\textcolor{bluetext}{\textbf{while}}}
\algrenewcommand\algorithmicif{\textcolor{bluetext}{\textbf{if}}}
\algrenewcommand\algorithmicelse{\textcolor{bluetext}{\textbf{else}}}
\algrenewcommand\algorithmicfor{\textcolor{bluetext}{\textbf{for}}}
\algrenewcommand\algorithmicthen{\textcolor{bluetext}{\textbf{then}}}
\algrenewcommand\algorithmicdo{\textcolor{bluetext}{\textbf{do}}}
\algrenewcommand\algorithmicreturn{\textcolor{bluetext}{\textbf{return}}}
\algrenewcommand\algorithmicend{\textcolor{bluetext}{\textbf{break}}}
\algrenewcommand\algorithmicend{\textcolor{bluetext}{\textbf{end}}}

\newtcolorbox{promptbox}[3][]{
arc=3mm,
lower separated=false,
fonttitle=\bfseries,
fontupper=\footnotesize,
colbacktitle=green!10,
coltitle=green!50!black,enhanced,
attach boxed title to top left={xshift=0.5cm,yshift=-2mm},
colframe=green!50!black,
colback=green!10,
title=#2,#1}

\newtcolorbox{responsebox}[3][]{
arc=3mm,
lower separated=false,
fonttitle=\bfseries,
fontupper=\footnotesize,
colbacktitle=red!10,
coltitle=red!50!black,
enhanced,
attach boxed title to top left={xshift=0.5cm,yshift=-2mm},
colframe=red!50!black,
colback=red!10,
title=#2,#1}

\newtcolorbox{keytakeaway}[3][]{
arc=1mm,
lower separated=false,
fonttitle=\bfseries,
colbacktitle=gray!10,
coltitle=green!30!black,
enhanced,
attach boxed title to top left={xshift=0.3cm,yshift=-2mm},
colframe=red!20!black,
colback=white, 
title=#2,#1}

\newcommand{\redcolor}{\color{black}}  
\newcommand{\ourapproach}{\textsc{VLM-Fuzz}\xspace}

\newcommand{\deepgui}{\emph{DeepGUI}\xspace}
\newcommand{\ape}{\emph{APE}\xspace}
\newcommand{\numapps}{59\xspace}

\newcommand{\sampleapps}{6\xspace}

\newcommand{\usedvision}{46\xspace}
\newcommand{\novision}{13\xspace}
\newcommand{\ablation}{5\xspace}

\newcommand{\classcov}{68.5\xspace}
\newcommand{\methodcov}{53.2\xspace}
\newcommand{\linecov}{46.5\xspace}
\newcommand{\deepclasscov}{57.8\xspace}
\newcommand{\deepmethodcov}{48.0\xspace}
\newcommand{\deeplinecov}{43.7\xspace}
\newcommand{\apeclasscov}{59.5\xspace}
\newcommand{\apemethodcov}{49.5\xspace}
\newcommand{\apelinecov}{44.4\xspace}

\newcommand{\aclasscov}{9.0\%\xspace}
\newcommand{\amethodcov}{3.7\%\xspace}
\newcommand{\alinecov}{2.1\%\xspace}

\newcommand{\dclasscov}{10.7\%\xspace}
\newcommand{\dmethodcov}{5.2\%\xspace}
\newcommand{\dlinecov}{2.8\%\xspace}

\newcommand{\novil}{5.28\%\xspace}

\newcommand{\realworldapps}{80\xspace}
\newcommand{\numvlmtest}{10\xspace}

\newcommand{\reportedbuggyapps}{50\xspace}
\newcommand{\realbuggyapps}{24\xspace}
\newcommand{\numcrashes}{208\xspace}
\newcommand{\exceptiontypes}{45\xspace}

\begin{document}

\title{\ourapproach: Vision Language Model Assisted Recursive Depth-first Search Exploration for Effective UI Testing of Android Apps}


\author{Biniam Fisseha Demissie \and
        Yan Naing Tun \and 
        Lwin Khin Shar \and 
        Mariano Ceccato
}


\institute{Biniam Fisseha Demissie \at
 Technology Innovation Institute \\
 9639 Masdar City, Abu Dhabi, UAE\\
\email{biniam@ymail.com}
           \and
Yan Naing Tun \at
 Singapore Management University \\
 Singapore, Singapore\\
\email{yannaingtun@smu.edu.sg}
            \and
Lwin Khin Shar \at
 Singapore Management University \\
 Singapore, Singapore\\
\email{lkshar@smu.edu.sg}
            \and
Mariano Ceccato \at
 University of Verona \\
 Verona, Italy\\
\email{mariano.ceccato@univr.it}            
}

\date{Received: date / Accepted: date}

\maketitle

\begin{abstract}
    Testing Android apps effectively requires a systematic exploration of the app's possible states by simulating user interactions and system events. While existing approaches have proposed several fuzzing techniques to generate various text inputs and trigger user and system events for UI state exploration, achieving high code coverage remains a significant challenge in Android app testing. 
    The main challenges are (1) reasoning about the complex and dynamic layout of UI screens; (2) generating required inputs/events to deal with certain widgets like pop-ups; and (3) coordination between current test inputs and previous inputs to avoid getting stuck in the same UI screen without improving test coverage. 
    To address these problems, we propose a novel, automated fuzzing approach called \ourapproach for effective UI testing of Android apps. We present a novel heuristic-based depth-first search (DFS) exploration algorithm, assisted with a vision language model (VLM), to effectively explore the UI states of the app. We use static analysis to analyze the Android Manifest file and the runtime UI hierarchy XML to extract the list of components, intent-filters and interactive UI widgets. VLM is used to reason about complex UI layout and widgets on an on-demand basis. Based on the inputs from static analysis, VLM, and the current UI state, we use some heuristics to deal with the above-mentioned challenges. 
    We evaluated \ourapproach based on a benchmark containing \numapps apps obtained from a recent work and compared it against two state-of-the-art approaches: \ape and \deepgui. \ourapproach outperforms the best baseline by \aclasscov, \amethodcov, and \alinecov in terms of class coverage, method coverage, and line coverage, respectively. 
    \redcolor{We also ran \ourapproach on \realworldapps recent Google Play apps (i.e., updated in 2024). \ourapproach detected \numcrashes unique crashes in \realbuggyapps apps, which have been reported to respective developers.} 
\end{abstract}

\keywords{Android, Fuzzing, GUI testing, Large Language Model, Vision Language Model}




\section{Introduction}
\label{sec:introduction}

Mobile apps have become an integral part of our daily lives. While developers must ensure app quality, they often work under time-to-market pressure, primarily focusing their testing on core functionalities and common usage scenarios. This often leads to released apps containing bugs or vulnerabilities which may severely impact user experiences and thus, the app's popularity.  
To deal with inadequate testing, automatic Android fuzzing has been an active area of research in software engineering community for the past decade. 

Existing GUI testing approaches can be categorized based on their exploration strategies such as random~\cite{monkey}, search-based~\cite{guiripper-ase12,amalfitano2014mobiguitar, azim2013targeted, mao2016sapienz}, model-based~\cite{su2017guided,dong2020time}, or program analysis based~\cite{anand2012automated,moran2017crashscope} approaches. But these approaches still have limitations in terms of (1) dealing with the evolving nature of Android platform; (2) scalability of the test generators when dealing with a large state space; and (3) low code coverage when dealing with complex interactions with the user, the system, and other apps because test inputs generated by these approaches do not resemble real users' actions~\cite{peng2022mubot}. To deal with these challenges, some approaches have incorporated machine learning techniques such as deep learning~\cite{yazdanibanafshedaragh2021deep} and reinforcement learning~\cite{pan2020reinforcement, romdhana2022deep}. However, deep learning or reinforcement learning approaches generally require large amount of labeled training data and it may not be scalable to label a representative list of GUI items.

Large Language Models (LLMs) such as GPT and Gemini have emerged as a powerful tool for  natural language understanding and image recognition. Recent advances in LLM have triggered various studies examining the use of these models for software engineering tasks such as code completion~\cite{du2024evaluating} and test generation~\cite{ryan2024code}. 
As such, recent approaches have applied large language models for GUI testing of Android apps~\cite{10.1145/3597503.3608137,liu2024make}. These approaches essentially formulate the GUI testing problem as a questions \& answering task, i.e., asking the LLM to play the role of a human tester to test the target app, with various text prompting techniques. While these approaches may be effective, by leveraging only text-based prompting techniques, they have not explored the full capability of LLMs where they can be leveraged to reason with images such as screenshots of activity component screens. They have also not explored the synergistic combination of search/model-based state exploration strategies with LLMs. While LLMs may be good at generating test inputs for a given GUI, a good state space exploration strategy is still needed to keep track of explored and unexplored states and to identify promising paths that could lead to better code coverage.

Our goal here is to address the low code coverage limitation highlighted in~\cite{peng2022mubot}. The main challenges we aim to address in this paper are \encircle{1}  reasoning about complex and dynamic layout of UI screens; \encircle{2}  generating required inputs/events to deal with certain widgets like pop-ups; and \encircle{3}  coordination between current test inputs and previous inputs to avoid getting stuck in the same UI screen without improving test coverage. 

We propose \ourapproach, a novel automated fuzzing approach for effective UI testing of Android apps. The approach first analyzes the AndroidManifest file to identify app components and relevant system events for each component, and it examines the dynamic UI hierarchy XML to capture interactive UI widgets. To optimize testing efficiency, \ourapproach evaluates component complexity based on the number of interactive UI elements and allocates testing time budget accordingly.
It then generates appropriate inputs/events to different widgets leveraging both heuristics and Vision Language Model (VLM). It adopts a heuristic-based recursive depth-first search (DFS) exploration strategy to explore different UI states of the app.  

The contributions of this paper are 
\begin{enumerate}
    \item \emph{Novel Android app fuzzing algorithm}: We propose an algorithm that combines a heuristic-based recursive depth-first search exploration strategy with Vision Language Model, which addresses the challenges involved in reasoning with complex, dynamic UI layouts and exploring the state space in an efficient manner.

    \item \emph{Comparative study with state of the arts (SOTAs)}: we compare our approach against  \ape~\cite{gu2019practical}  and \deepgui~\cite{yazdanibanafshedaragh2021deep} --- based on a benchmark consisting of \numapps apps. Our approach achieves a high code coverage (\classcov\% class coverage, \methodcov\% method coverage, and \linecov\% line coverage  on average), outperforming the best baseline by \aclasscov, \amethodcov, and \alinecov in terms of class coverage, method coverage, and line coverage, respectively. 
    \redcolor{We also compare our approach against GPTDroid~\cite{liu2024make}, a recent and closely related tool, \emph{analytically} (as its tool was unavailable despite our requests). We were able to replicate it following the steps given in the article, which are essentially iterative promptings of GUI information.} 

    \item \emph{Ablation study on the usefulness of vision LM}: We investigate the  VLM's ability to reason with GUI objects and how useful it is in improving the test coverage.

    \item \redcolor{\emph{Bug detection capability on real world apps}: We further evaluate our approach on \realworldapps recent official apps from Google Play. \ourapproach induced crashes in \reportedbuggyapps apps, among which we manually confirmed that \numcrashes crashes in \realbuggyapps apps are due to real bugs. The results show that \ourapproach is effective in testing real-world apps.}

\end{enumerate}

The rest of the paper is organised as follows. Section~\ref{sec:background} discusses some background on Android fuzzing and specific challenges that motivate our work. Section~\ref{sec:overview} provides an overview of our approach. Section~\ref{sec:approach} presents the details of our approach. Section~\ref{sec:evaluation} evaluates our approach. Section~\ref{sec:related} discusses related works.
Section~\ref{sec:conclusion} concludes the paper and discusses future work.

\section{Background}
\label{sec:background}

This section introduces relevant background on Android UI fuzzing and its challenges.

\subsection{GUI of Android Apps}
In Android, an activity component represents a single screen with a user interface. It is the entry point for interacting with the user.
The UI is usually defined in the XML layout files located inside a specific folder of the APK. When an activity component is activated, e.g., when an app launch event occurs, the XML layout files are parsed and converted into a tree of ``View'' objects.
``View'' is the base class for all UI components (e.g., buttons, text fields). Each view object renders a rectangular area on the screen. Widgets are a specific type of View objects, which is a UI component used to interact with the user.  
Common examples of widgets include buttons, text fields, checkboxes, and image views. It supports various GUI actions such as clicking and swiping. A widget has four categories of attributes describing its type (e.g., class), appearance (e.g., text), functionalities (e.g., clickable and scrollable), and the designated order among sibling widgets (i.e., index). Each attribute is a key-value pair. 

\begin{figure}[h]
    \centerline{\includegraphics[width=0.30\textwidth]{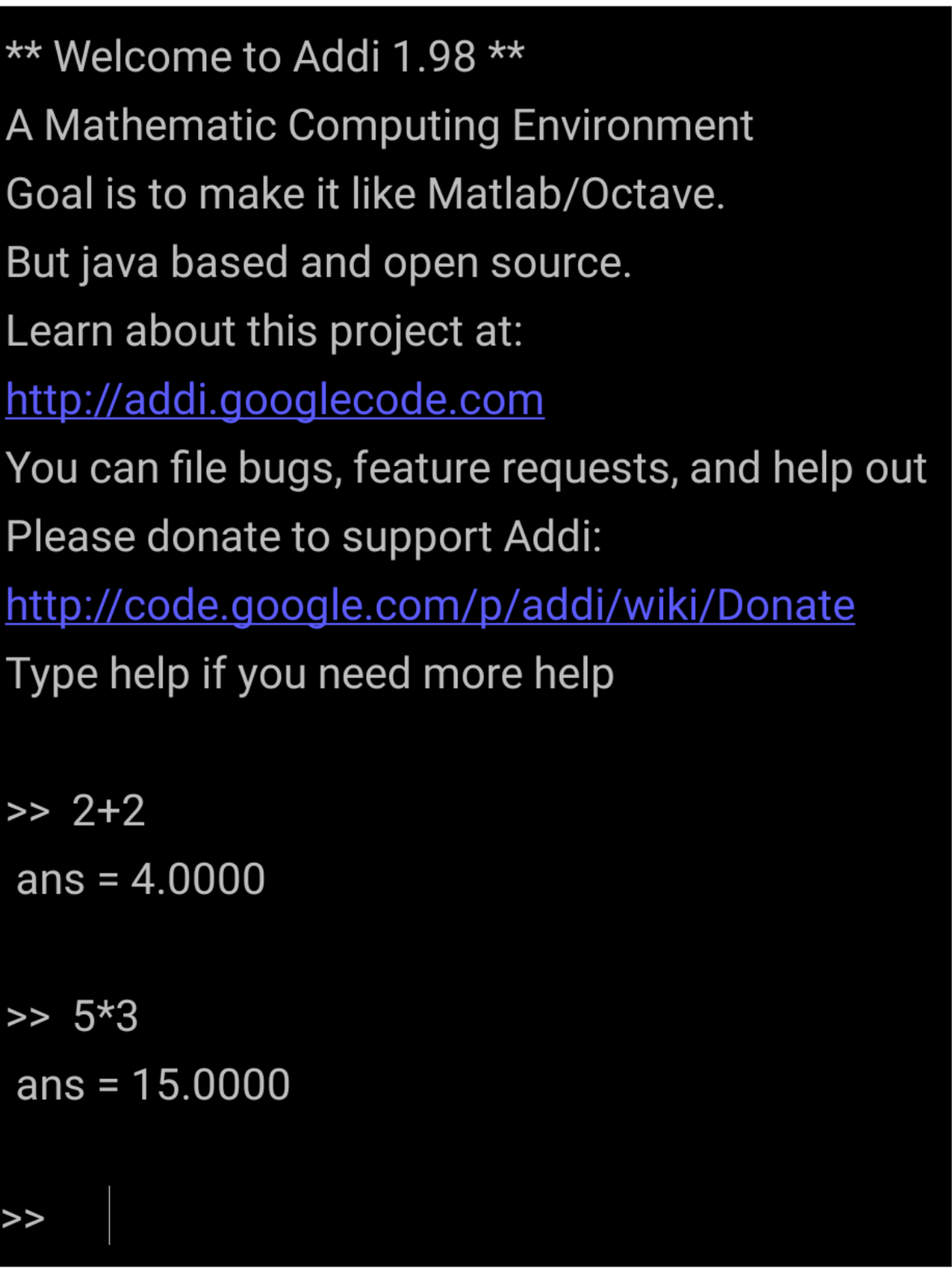}}
  \caption{An example of UI challenging for typical UI fuzzers (shown with \ourapproach generated input).}\label{fig:complexui}
  \vspace*{-0.8em}
\end{figure}

\begin{figure}[bh]
    \centering 
    \includegraphics[width=0.99\textwidth]{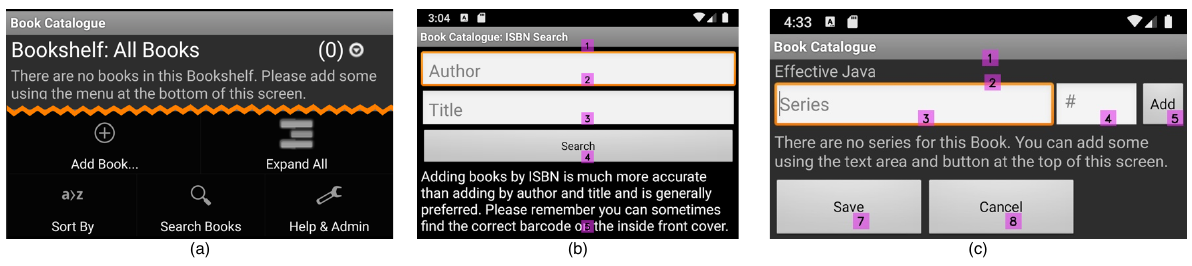}
    
    \caption{Example, complex UIs showing (a) a UI overlaid on top of another UI; (b) a UI where complex, valid input is required; (c) a UI where a particular sequence of events is required.}
    \label{fig:runexample}
\end{figure}

It is important for UI fuzzers to be able to reason about complex UI layout and different ``View'' objects, and their context, to generate valid inputs.  As an example, ~\autoref{fig:complexui} shows an activity screen that requires reasoning about the context of the app screen and the meaning of symbols such as \texttt{>>}, which would be challenging for typical UI fuzzers. Without any form of intelligence and reasoning capability, the fuzzer is unlikely to automatically detect that this is a calculator screen and that the input fields following the symbols \texttt{>>} require numeric inputs. 
\autoref{fig:runexample}.a shows an activity screen where a dynamic pop-up menu is overlaid on top of an initial UI. In this case, a typical depth first search strategy may simply continue to explore on the pop-up menu while there may still be remaining states left unexplored on the initial UI screen.
\autoref{fig:runexample}.b shows a UI in which reasoning of the UI context is required to generate valid inputs such as ISBN number. \autoref{fig:runexample}.c shows a UI where a particular sequence of events is required. This shows that a fuzzer is required to have both reasoning capability of the UI context and a good state space exploration strategy to handle such complex UIs, which are common in modern apps nowadays.

\subsection{Exploration Strategy}
\emph{Random search} is a simple yet often effective method where UI events are generated randomly. It is employed by the widely-used Monkey test tool~\cite{monkey}. Though this strategy is useful when there is little knowledge about the input domain, often inefficient in terms of coverage, as it may repeatedly test similar or irrelevant inputs. 
\emph{Depth First Search (DFS)} is a systematic method for exploring the state space by diving deep into each branch before backtracking. This strategy is used in Android fuzzers like GUIRipper~\cite{guiripper-ase12}, MobiGUITAR~\cite{amalfitano2014mobiguitar} and A3E~\cite{azim2013targeted}. This strategy can also be inefficient without proper handling since DFS can get stuck in cyclic paths or in deep branches.
\emph{Evolutionary-based search} strategies, such as Genetic Algorithms (GA), simulate the process of natural evolution to optimize test generation. This strategy is used in Prev~\cite{demissie2020security}. Although evolutionary search can be powerful in many contexts due to its adaptive nature in exploring the state space, 
it is often challenging to define appropriate fitness function for GUI testing. GUI interactions might lead to subtle changes in the app's state that are hard to capture in a fitness function. For instance, a UI screen might appear to be the same as previous test outputs and hence, a fitness function may evaluate that this screen has been fully tested but it may still have subtle states that may not be captured by fitness functions.
\emph{Model-based search} strategies, such as Stoat~\cite{su2017guided} and \ape~\cite{choi2018practical} generate app event sequences according to models extracted from project artifacts such as source code, XML configuration files, and UI runtime state. These approaches enable the representation of app behavior as a model, allowing for the application of various exploration strategies. However, complex widgets like dynamic pop-ups (e.g.,~\autoref{fig:runexample}.a), commonly used in apps, present challenges in model construction, often resulting in an incomplete model. 

These observations motivate us to propose a hybrid approach that combines DFS and random exploration strategies, augmented with heuristics to avoid getting stuck in cyclic paths and with vision-based reasoning of GUIs. Overall, our approach is designed to overcome the challenges involved in effective exploration of complex GUIs.

\section{Overview}
\label{sec:overview}

\begin{figure}[tbh]
\centering
\includegraphics[trim={0.5cm 0.2cm 0 0.3cm},clip,width=1\textwidth]{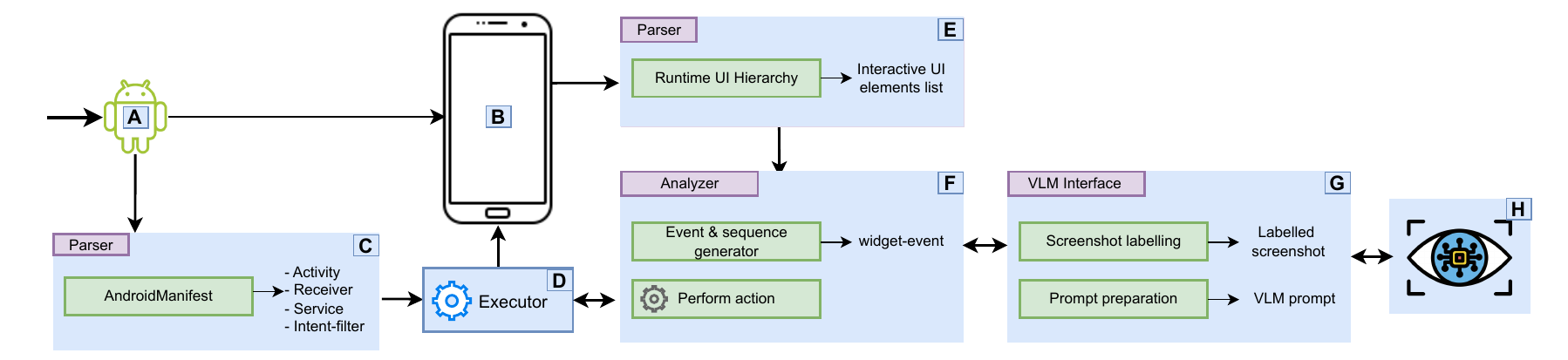}
\caption{Approach overview}
\label{fig:overview}
\end{figure}

\autoref{fig:overview} shows an overview of \ourapproach.
Given the path of an APK file (A) as an input, \ourapproach
parses the AndroidManifest file (C) and obtains the list of components (activities, services, and broadcast receivers) along with their intent-filters --- a section that declares the capabilities of a given component. 
From this list, \ourapproach randomly selects a component and instantiates it on the emulator/device (B). If the component is visible, \ourapproach identifies and extracts the list of interactive UI items (E) from the current UI hierarchy dumped using a custom-made tool that leverages the Accessibility Service. The analyzer (F) then generates the appropriate events to the different widgets using different heuristics leveraging VLM (G-H). Generated events are then sent to the UI widgets by the executor (D).

The following section explains the approach in detail.

\section{Approach}
\label{sec:approach}
{\redcolor Our approach offers a novel method for fully automated Android app testing by integrating AI-driven semantic understanding (using a Vision Language Model) with a traditional Depth-First Search (DFS) algorithm for state exploration.  Unlike existing approaches that often rely on model-based or coverage-guided techniques, this method emulates human-like understanding of UI interactions and emphasizes thorough state traversal, setting it apart from other approaches.
}

Algorithm~\ref{alg:executor} describes our main Executor algorithm. Our approach begins with a preliminary computation of component budgets. Instead of distributing the budget equally among components, we base our calculations on the number of interactive widgets and menu items present in each UI. Some simple UIs, despite their apparent straightforwardness, could potentially trap the recursive testing approach in infinite loops. For instance, each interaction might add a new item to a ListView without improving overall coverage, leaving complex UIs with insufficient testing time. This preliminary assessment is conducted by launching each component and enumerating the distinct {\em interactive} widgets and menu items present. The APK's AndroidManifest file is dumped using AAPT\footnote{https://developer.android.com/tools/aapt2}, and we use a custom-made parser to extract app components and their intent-filters. Then, each component is systematically launched by sending explicit Intents that satisfy the intent-filters (function {\em extractComponents()}). Upon successful component launch, the component's UI hierarchy XML is retrieved using a custom-made UI hierarchy XML dumper that leverages Accessibility Service\footnote{https://developer.android.com/reference/android/accessibilityservice/AccessibilityService} 
and the component budget is computed (function {\em getComponentsBudget()}). This approach ensures that more complex components with several interactive UI elements receive proportionally more testing time, while simpler components with fewer UI elements are allocated less time, ensuring comprehensive coverage across varying UI complexities.


\begin{sansserifalgoritm}
\caption{Executor()}
\begin{algorithmic}[1] 
\small
\Statex
\Statex \textcolor{bluetext}{\textbf{Input:}} Path to the APK to be tested ($apkPath$).
\Statex \hspace{0.9cm} Allocated budget (mins) for the entire test ($budget$).
\Function{Executor}{$apkPath, budget$}

    \State $componentsList \gets \text{extractComponents}()$
    \State $componentBudgetList \gets \text{getComponentsBudget}(componentsList, buudget)$
    \For{$component \in componentsList$} 

        \State $startComponent(component)$

        \State $budget \gets componentBudgetList[component]$
        
        \State ${\bf UNTIL(budget)}:UI\_Analyzer(component, UIStack, nonIgnoreComponentList)$
        
    \EndFor

\EndFunction

\end{algorithmic}
\label{alg:executor}
\end{sansserifalgoritm}


Once the proportional budget is computed, the algorithm  explores the Android app's UI state space. It uses a heuristic-based recursive depth-first search. Similar to the preliminary assessment phase, it launches each component by sending explicit Intents and dumps the corresponding UI hierarchy (function {\em startComponent()}). {\redcolor We start each component via explicit Intent to ensure reliable testing, as components may not be directly accessible from the main Activity. Additionally, components can exhibit different behaviors based on the Intent's data and action, allowing us to test various scenarios independently.} The algorithm then parses the XML output to identify interactive widgets on the currently visible portion of the component. A mapping of widget to action is then generated for each identified interactive widget (e.g., a tap action for a button). The subsequent step involves interacting with these items while monitoring for UI state change. The recursive process is repeated if UI state change is detected. A UI state change event is fired when, (i) on the same component, new widgets appear (e.g., after a scroll action or a new list item is added), (ii) a popup overlay is shown on the same component (\autoref{fig:runexample}.a), or (iii) a component switch occurs. The component's UI is then explored until its budget is exhausted (function {\em UI\_Analizer()}). 

 We developed a recursive UI state explorer that leverages Vision Language Models (VLMs) on complex UIs. VLMs integrate computer vision with natural language processing, enabling them to interpret images and text together. This dual capability allows them to handle tasks requiring both visual understanding and linguistic comprehension. In this work, we use OpenAI's GPT-4o Vision, a state-of-the-art multimodal AI model~\cite{gpt4vision}. Before selecting GPT-4o, we did a preliminary experiment with several VLMs, including GPT-4o, Qwen2-VL-7b, Gemini-1.5, Pixtral-12b, and Claude 3.5 Sonnet, via HuggingFace's VisionArena~\cite{vision-arena}. In our preliminary experiment, the VLMs were presented with \numvlmtest screenshots of UIs randomly extracted from our benchmark apps and were prompted to generate a sequence of events to exercise the presented UIs using the prompt shown in~\autoref{fig:prompt-response}. The default configuration of the VLMs was used. One of the authors manually verified the VLMs' outputs. GPT-4o achieved the most accurate results. {\redcolor GPT-4o likely performed better due to its enhanced ability to process visual information from UI screenshots and understand spatial relationships between interface elements, combined with its strong capabilities in generating sequential instructions and understanding UI interaction patterns.} As an example, for the UI screen given in~\autoref{fig:runexample}.c, the sequence of events generated by a few VLMs are shown in~\autoref{tab:vlmtest} where GPT-4o produced the most accurate sequence of events. 
 We made all these experiment results --- the sequence of events generated by various VLMs for the 10 UI screenshots  --- available in our replication package~\cite{vlm-fuzz}. 
 
 \begin{table}[tbh]
\caption{Sample sequence of events generated by VLMs for the UI screen in~\autoref{fig:runexample}.c}
\centering
 \footnotesize
  \begin{tabular}{lll}
 
   &   \textbf{VLM} & \textbf{Sequence of Events}    \\ \hline

   Commercial &   GPT-4o & \makecell[l]{[tap(3); input(3, "Series 1"); tap(4); input(4, "1"); \\tap(5); tap(7);]} \\

    Commercial & Claude3.5 Sonnet & [input(2, "Harry Potter"); input(4, "1"); tap(5); tap(7);] \\

   Commercial &  Gemini-1.5-flash &  \makecell[l]{[tap(2); input(2, "The Lord of the Rings"); tap(4); \\input(3, "The Hobbit"); tap(7);]} \\
    \hline
    
    Open source & Qwen2-VL-7B & [tap(2); input(2, \textbackslash"Series Name\textbackslash");] \\

    Open source & minicpm-llama3-v & [tap(3); input(3, \textbackslash"The Lord of the Rings\textbackslash"); tap(5);] \\

    Open source & pixtral & [tap(3); input(3, "Harry Potter"); tap(5);] \\
    \hline
    \end{tabular}
\label{tab:vlmtest}
\end{table}

Below, we describe the UI analyzer.

\subsection{UI Analyzer}
\label{sec:uianalyzer}
Algorithm~\ref{alg:ui-analyzer} outlines our UI Analyzer approach. The UI analyzer is the core part of our approach that recursively analyzes the current UI, dumping its UI hierarchy, extracting interactive UI widgets, performing interactions, and monitoring UI state changes. {\redcolor It takes three inputs: the currently visible component's details, a stack tracking previously visited components, and a list of relevant external components (such as Android framework dialogs for permission requests) that, while not part of the app under test, are critical to restore app functionality.} Moreover, the algorithm implements several heuristics to maximize new state discovery. We set a threshold $\tau$ to limit the number of times a UI component can be analyzed, where our experiments showed that $\tau = 2$ achieves good test coverage while staying within budget (60 minutes). This means once a component has been analyzed twice, any further state changes in that UI element will not trigger a new analysis. When a component is ready for analysis and contains a text editor widget (e.g., {\em EditText}), the algorithm first tries to use the VLM-assisted action performer (via the {\em performVisionActions()} function). {\redcolor The VLM is prioritized because it can both generate contextually relevant text input and determine appropriate sequences of UI events to accomplish tasks presented on the screen.} If this function successfully explores all widgets, the test is considered complete. {\redcolor Considering that the VLM's output may sometimes be ambiguous or imperfect, we have implemented a robust fallback mechanism. This mechanism cross-checks the suggested actions against the static properties of the UI elements. If the VLM-assisted approach fails to achieve complete UI coverage or encounters screens without text editors, the algorithm falls back to the non-VLM assisted action performer (via the {\em performNonVisionActions()} function), which provides comparable exploration effectiveness for non-text interactions while being cheap costwise. Functions {\em performVisionActions()} and {\em performNonVisionActions()} are described in the~\autoref{sec:appendix}.
}


\begin{sansserifalgoritm}
\caption{UI Analyzer}
\begin{algorithmic}[1]
\small
\Statex \textcolor{bluetext}{\textbf{Input:}} Current visible component ($currentComponent$).
\Statex \hspace{0.9cm} UI component visit stack ($UIStack$).
\Statex \hspace{0.9cm} List of non-AUT\footnotemark components not to ignore  ($nonIgnoreComponentList$).
\Function{UI\_Analyzer}{$currentComponent, UIStack, nonIgnoreComponentList$}
    \State $currentComponentDetails \gets \text{getCurrentComponentDetails}(currentComponent)$

    \If{$\text{visitCount}(currentComponent) > \tau $}
        \State \Return
    \EndIf

    \If{$\text{currentComponent} \in UIStack $}
        \If{$\text{\textbf{\textcolor{bluetext}{not }}}\text{UIItemsChanged}()$}
            \State \Return
        \EndIf
    \EndIf    

    \If{$currentComponent \text{\textbf{\textcolor{bluetext}{ not}}} \in AUT$}
        \State $\text{press\_back}()$        
        \State \Return
    \EndIf        

    \If{$currentComponent \text{\textbf{\textcolor{bluetext}{ not}}} \in UIStack$}
        \State $\text{UIStack.push}(currentComponent)$        
    \ElsIf {$UIStack[lastEntry] == currentComponent$}
        \State \Return
    \EndIf            

    \State $widgetsList \gets \text{currentComponent.getWidgets}()$
    \If{$\text{\textbf{\textcolor{bluetext}{not }}}\text{performVisionActions}(widgetsList)$}
        \State $\text{performNonVisionActions}(widgetsList))$        
    \EndIf      

    \State $\text{UIStack.pop}()$ 
    \State \Return 
\EndFunction
\end{algorithmic}
\label{alg:ui-analyzer}
\end{sansserifalgoritm}
\footnotetext{App Under Test}


\subsubsection{\bf{UI with text input widgets}}
\label{sec:vlm}
\begin{figure}[!tbh]
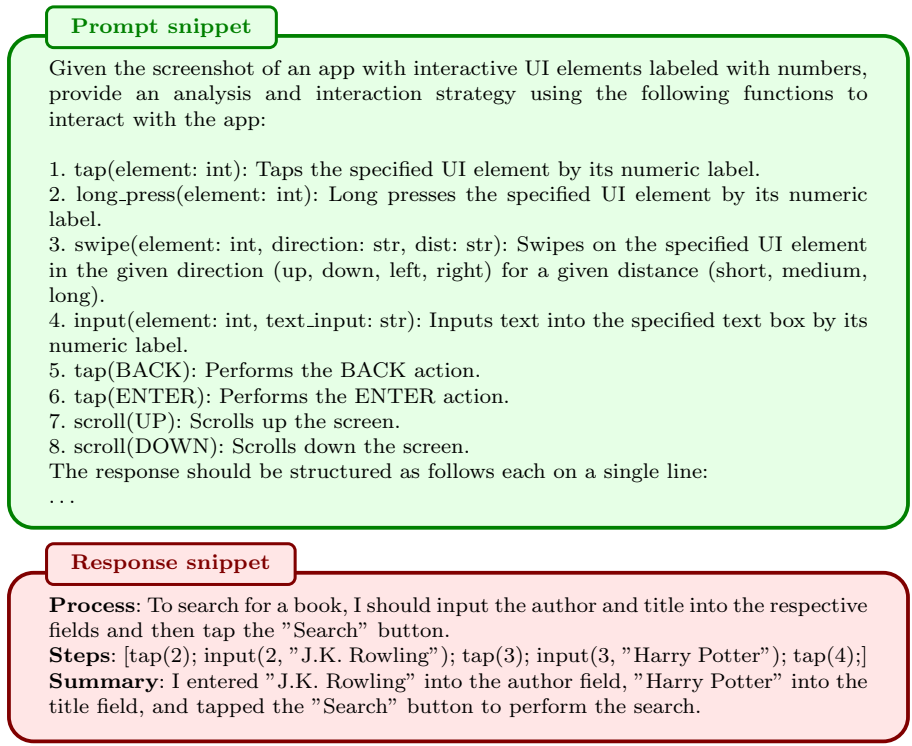

\begin{promptbox}[]{Prompt snippet}
\verbatim
 Given the screenshot of an app with interactive UI elements labeled with numbers, provide an analysis and interaction strategy using the following functions to interact with the app:\\

1. tap(element: int): Taps the specified UI element by its numeric label.\\
2. long\_press(element: int): Long presses the specified UI element by its numeric label.\\
3. swipe(element: int, direction: str, dist: str): Swipes on the specified UI element in the given direction (up, down, left, right) for a given distance (short, medium, long).\\
4. input(element: int, text\_input: str): Inputs text into the specified text box by its numeric label.\\
5. tap(BACK): Performs the BACK action.\\
6. tap(ENTER): Performs the ENTER action.\\
7. scroll(UP): Scrolls up the screen.\\
8. scroll(DOWN): Scrolls down the screen.\\
The response should be structured as follows each on a single line:\\
\dots
\end{promptbox}

\begin{responsebox}[]{Response snippet}
\verbatim
{\bf Process}: To search for a book, I should input the author and title into the respective fields and then tap the "Search" button.\\
{\bf Steps}: [tap(2); input(2, "J.K. Rowling"); tap(3); input(3, "Harry Potter"); tap(4);]\\
{\bf Summary}: I entered "J.K. Rowling" into the author field, "Harry Potter" into the title field, and tapped the "Search" button to perform the search.
\end{responsebox}
\caption{Example prompt snippet and VLM response}
\label{fig:prompt-response}
\vspace{-1em}
\end{figure}

If the UI contains text widgets, a screenshot is taken and each interactive widget is labelled with numbers (e.g., ~\autoref{fig:runexample}.b). The labelled screenshot along with a one-shot prompt is then fed to the VLM. The structured output is parsed and actions are mapped to their corresponding labeled widgets, and the actions are performed following the sequences suggested by the VLM. In case of a text input event, the generated text is sent to the target widget, and any resulting state changes on the widget are monitored. If no state changes occur, for example, because alphanumeric text was generated when only numeric input is accepted (e.g., widget 4 on~\autoref{fig:runexample}.c), an alternative input (e.g., a numeric value) is generated and sent. {\redcolor Since datatype is unknown at runtime, this} verification is performed by comparing the {\em text} property of the current widget with the generated text. A mismatch indicates that the generated text was rejected. ~\autoref{fig:prompt-response} shows a snippet of a prompt and the corresponding VLM response. {\redcolor The output of the VLM is sensitive to the precise wording of the prompt and its configuration. To address this, we adopted an iterative refinement process, starting with a basic instruction to generate event sequences for a given UI screenshots to standardize the prompt structure, ensuring consistent outputs. Finally, we leveraged GPT-4o to reverse-engineer and formalize the optimal prompt structure that consistently produced the desired output format and quality.} In  this example, we used the UI in \autoref{fig:runexample}.b where author name and title of a book by the author is expected to be filled in before tapping the {\em Search} button. The VLM generates the steps and inputs necessary to complete the tasks on that UI.

If there are any unprocessed widgets remaining after performing the VLM suggested actions (e.g., the VLM missed an action for an interactive widget), a non-VLM based state exploration is launched. If any of the unprocessed widgets is a text input box, attributes associated to the widget (e.g., widget ID {\em "authorname"}, or placeholder text {\em "Author"}) are extracted and structured as a JSON object and fed into an LLM to predict an appropriate text input. If this fails, a random text or number is generated. The remaining actions are inferred from the UI hierarchy and performed following the heuristic described in Section~\ref{subsec:ui-without-text}.  

After all possible interactions are complete, the subsequent interactive operation is the Menu tap, a function that activates the menu if the UI implements one. An approach solely based on VLM/LLM would miss the exploration of such hidden interactions as in some cases there is no indicator of existence of menus either in the visible UI or in the UI hierarchy XML. \ourapproach taps on the Menu button to see whether the app implements menus and monitors UI change. If a UI change occurs, i.e., a pop-up menu is overlaid on the current screen (see~\autoref{fig:runexample}.a), the recursive process restarts for the current UI. {\redcolor While traditional Menu-based navigation is less common in modern Android applications, our testing approach must remain comprehensive to ensure fair comparisons with existing tools. The state-of-the-art testing frameworks include Menu interaction capabilities, and excluding Menu testing would compromise VLM-FUZZ's performance evaluation on benchmark applications that utilize these elements.}

~\autoref{fig:runexample}.c shows an example UI that requires a specific sequence of actions that might be difficult to discover through random exploration alone. While traditional heuristics could help determine the correct order of interactions, a VLM would be particularly well-suited for this task. In this specific example, the VLM successfully identified the correct sequence of user actions: {\em  [tap(3); input(3, "Java Series"); tap(4); input(4, "1"); tap(5); tap(7);]}.

Note that if a UI change occurs before all the possible states on that UI are explored, the events that led to the last state are replayed (except the last action that caused the UI change), and the remaining actions are performed. If replaying does not bring the state back, the UI is considered permanently altered and the exploration is ended for that UI. This process is described in details in Section~\ref{subsec:transition-record}.

\subsubsection{\bf{UI without text input widgets}}
\label{subsec:ui-without-text}
In the cases where no text input is necessary (e.g.,~\autoref{fig:runexample}.a), we perform the actions based on the following orders. First, tappable actions are performed followed by scroll actions. If the component implements a menu, the interaction with menu items is followed. The following heuristics are applied in order to maximize the new state discovery.

\paragraph{Tappable widgets:} Tappable widgets are grouped into three based on their text sentiment, namely, (i) neutral, (ii) positive (e.g., "Save" or "Open") or (iii) negative (e.g., "Exit", "Back" or "Cancel"). The rationale behind this is to perform the actions that could potentially lead to navigation away from the current UI as the last step, ensuring the majority of the actions are completed first. Each group is shuffled and interactions are performed in the order (i) $\rightarrow$ (ii) $\rightarrow$ (iii). Moreover, if there are tappable widgets appearing on the same Y-axis of the screen (e.g., {\bf Yes $|$ No $|$ Cancel}, similar to buttons 7 \& 8 in \autoref{fig:runexample}.c), the tap action on these widgets will be performed at the end of all other interactions. 

\begin{figure}[tbh]
\centering
\includegraphics[width=0.3\textwidth,trim={0 30cm 0 0},clip]{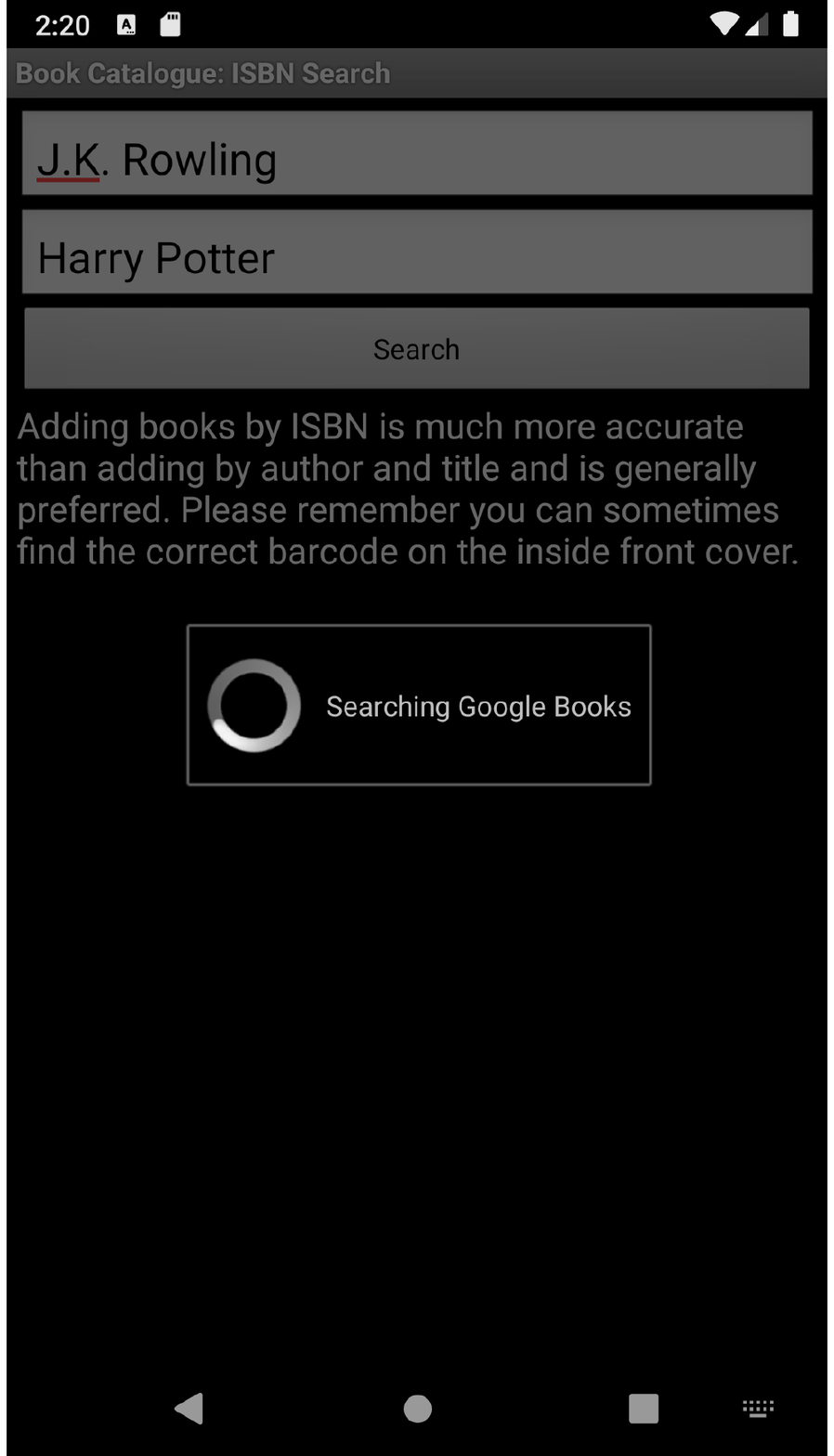}
\caption{A loading progress indicator}
\label{fig:loading-page}
\vspace{-0.5em}
\end{figure}

\paragraph{Progress bars:} After performing an action, most approaches wait for an idle state before evaluating the state change. In cases of progress indicators (e.g., a progress bar widget), though an idle state is reached (e.g., no new widgets are added to the UI within the idle period), an effective state exploration approach should wait until the progress bar finishes or a given timeout greater than idle period is reached. In fact, if a "{\em loading...}" indicator appears similar to the example shown in \autoref{fig:loading-page}, a human would naturally wait for it to finish before navigating away from this UI. Our approach works the same way. When it detects a visible progress indicator {\redcolor (via the XML UI hierarchy)}, it pauses the current analysis and monitors the UI for up to 60 seconds to detect any changes. If the UI changes (e.g., new UI or progress indicator is hidden) in this monitored period, a new instance of the analyzer is launched for the new state, otherwise the current analyzer is resumed.  This approach ensures we do not miss important UI states that appear after loading screens or progress indicators finish. In this example, once the search completes, a UI with a form opens where all the fields are already filled in with the search results.

\subsection{Transition Record}
\label{subsec:transition-record}
Typically, tap actions trigger UI changes by either displaying a popup overlay or navigating to a different screen. For instance, tapping the Menu button on the device might reveal a popup menu that appears on top of the current UI, as illustrated in~\autoref{fig:runexample}.a. Despite appearing as an overlay on the current UI, this popup maintains its own unique UI hierarchy XML, thus qualifying as a distinct UI state. The UI in~\autoref{fig:runexample}.a presents five interactive elements. Tapping on any of the items (e.g., \emph{Add Book...}), may navigate the analyzer away from the current view, possibly even exiting the app. In such cases, interaction with the remaining four widgets would be missed. To address this issue, \ourapproach records the exact sequence of state transitions, allowing it to replay these actions later to return to the same UI state and continue testing any remaining interactive elements that were not yet explored. In this context, a state $s \in S$, where $S$ is the set of all possible UI states of the app under test (AUT) is a tuple defined as follows:

\[ s = (W, P) \]

where:
\begin{itemize}
    \item $W$ is the set of UI elements (widgets) in that UI state
    \item $P$ is the set of properties for each element
\end{itemize}

A transition $T$ from state $s1$ to $s2$ is then defined as:

\[
   T: s1(W_1, P_1) \xrightarrow{a} \\ s2(W_2, P_2)
\]

where $a \in A$ is an action performed in state $s1$ and $A$ is the set of supported actions (See Section~\ref{sec:implementation}).





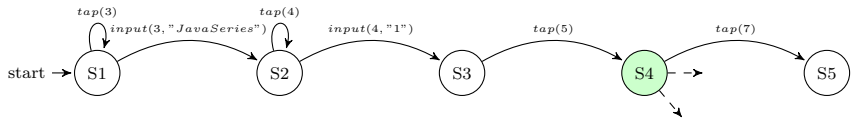
\begin{figure}[tbh]
\begin{tikzpicture}[->, >=stealth', shorten >=1pt, auto, node distance=3cm, scale=0.8, transform shape]

  \node[state, initial] (S1) {S1};
  \node[state, right of=S1] (S2) {S2};
  \node[state, right of=S2] (S3) {S3};
  \node[state, right of=S3, fill=green!20] (S4) {S4};
  \node[state, right of=S4] (S5) {S5};
  
  \path (S1) edge[bend left, above] node{\tiny $input(3, "Java Series")$} (S2)
        (S2) edge[bend left, above] node{\tiny $input(4, "1")$} (S3)
        (S3) edge[bend left, above] node{\tiny $tap(5)$} (S4)
        (S4) edge[bend left, above] node{\tiny $tap(7)$} (S5)
        (S1) edge[loop above] node{\tiny $tap(3)$} (S1)
        (S2) edge[loop above] node{\tiny $tap(4)$} (S2);
 \draw[->,dashed] (S4) -- node[sloped]  {\tiny } +(0:1);
 \draw[->,dashed] (S4) -- node[sloped,swap]  {\tiny } +(-50:1);
\end{tikzpicture}
\vspace{-0.4cm}
\caption{Example state transition}
\label{fig:transition}
\end{figure}

An example transition record is shown in~\autoref{fig:transition} for the UI in~\autoref{fig:runexample}.c where a component is started (S1) and a tap on widget 3 focuses the widget. A text input "JavaSeries" is inserted in widget 3 (S2) and widget 4 is tapped to focus. The text "1" is then inserted (S3) followed by taps on widget 5 and 7, respectively. If any subsequent tap action causes the app to change UI before all the interactive widgets are explored, once the current exploration finishes, the algorithm attempts to replay the transition record to come back to the last state, in this case $S4$ (backtracking), skipping the last action (i.e., tap(7)) that caused the change and explores the remaining interactive widgets.

\begin{figure}[tbh]
\centering
\includegraphics[width=0.8\textwidth]{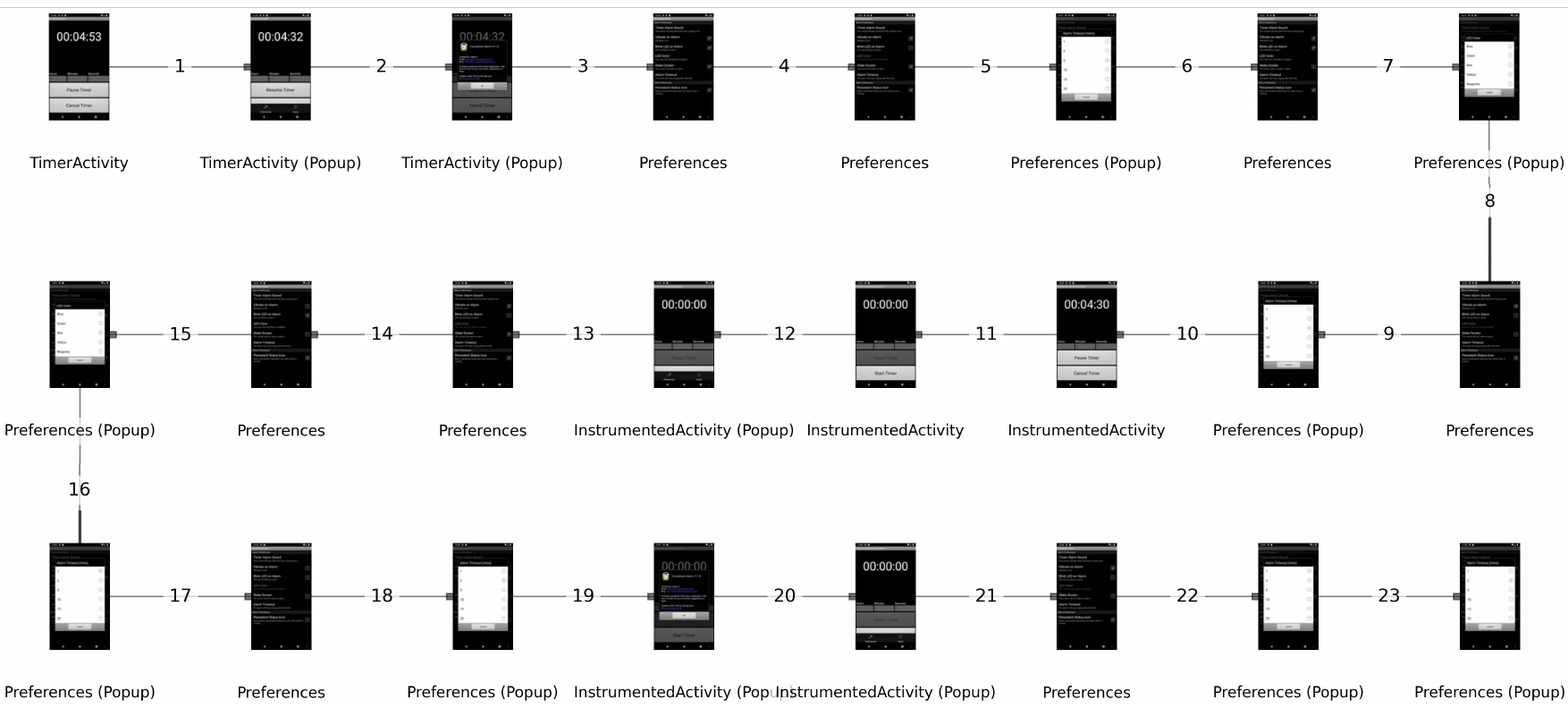}
\caption{Sample state space exploration of an app (\texttt{everythingandroid.timer})}
\label{fig:explore}
\end{figure}

As an example, \autoref{fig:explore} shows an excerpt of state space exploration of an app we experimented with. It shows that our approach was able to explore the possible states in the same activity screen despite the dynamic pop-ups (e.g., transitions 6 and 8) and escape from the current UI once a recursion without any new discovery state is detected (e.g., transitions 3 and 10).

\subsection{Implementation}
\label{sec:implementation}

\ourapproach is implemented using Python while the companion server app that is used to dump the UI hierarchy is developed in Java, and they communicate via broadcast. The server generates and dumps the UI layout hierarchy in a format and location identical to UIAutomator~\cite{uiautomator}, making it a suitable replacement for UIAutomator. \ourapproach runs both on emulators and real devices. It requires neither instrumentation nor modification to the app under test, Android device, or Android operating system. It only requires the device to be connected to the computer running \ourapproach tool via Android Debug Bridge (ADB). Apart from Android SDK tools such as ADB and Android Asset Packaging Tool (AAPT), \ourapproach does not have other dependencies.

\emph{Supported events}: \ourapproach supports generating the following events that can be simulated using ADB: 
\begin{enumerate}
    \item {\em UI events (actions)}: click, long-click, scroll, text-input, menu-tap, app switch and screen-rotate. VLM is used to generate these events.
    \item {\em Intents}: when launching components, \ourapproach uses the defined intent-filters and attempts to generate valid Intent. If there are multiple intent-filters, a random one is picked every time the component is launched.
    \item {\em System broadcast Intents}: valid broadcast Intents are generated based on the app's intent-filter definition. For example, if an app defines the \newline \texttt{TIMEZONE\_CHANGED} system broadcast Intent, the broadcast is sent to the app with \texttt{TIMEZONE} and \texttt{TIME\_PREF} extra fields populated. \ourapproach supports 187 system broadcast Intents. These system broadcast Intents are constructed once from the Android documentation\footnote{https://developer.android.com/about/versions/11/reference/broadcast-intents-30} and stored in file, where example inputs are extracted with the help of an LLM. Whenever a BroadcastReceiver component defines system broadcast Intent, the file is consulted to generate the right Intent (i.e., action, extras, etc) for the component. The file is publicly available in our tool repository to enable researchers and developers to extend existing fuzzers with system broadcast capabilities.
\end{enumerate}

Approaches developed on top of Android Monkey such as \ape can generate more UI events such as drag and pinch-zoom or volume level.
{\redcolor We recognize that this engineering limitation may restrict our tool's ability to fully exercise all app functionalities, potentially impacting bug detection in scenarios involving multi-touch interactions. In future work, we will extend our approach to incorporate a wider range of actions, including drag, pinch, and other multi-touch actions, to enhance test coverage and better simulate real-world user behavior.}

\section{Empirical Evaluation}
\label{sec:evaluation}

In this section, we present the design and results of our experiments to assess the efficacy of \ourapproach. 
Our experiments are designed to answer the following Research Questions~(RQs):

\begin{itemize}
    \item \textbf{RQ1. Code Coverage.} How effective is \ourapproach in terms of code coverage? How does \ourapproach compare against the state-of-the-arts, namely \deepgui and \ape? 
  
      \item \textbf{RQ2. Ablation Study.} How useful is VLM at determining relevant test inputs? And what is the cost of using VLM?
       \item \textbf{RQ3. Bugs Detection Capability.} How is the effectiveness of \ourapproach in detecting bugs in real-world apps?

\end{itemize}

\subsection{Benchmark Apps} 
\begin{figure}[tbh]
\centering
\includegraphics[width=0.6\textwidth]{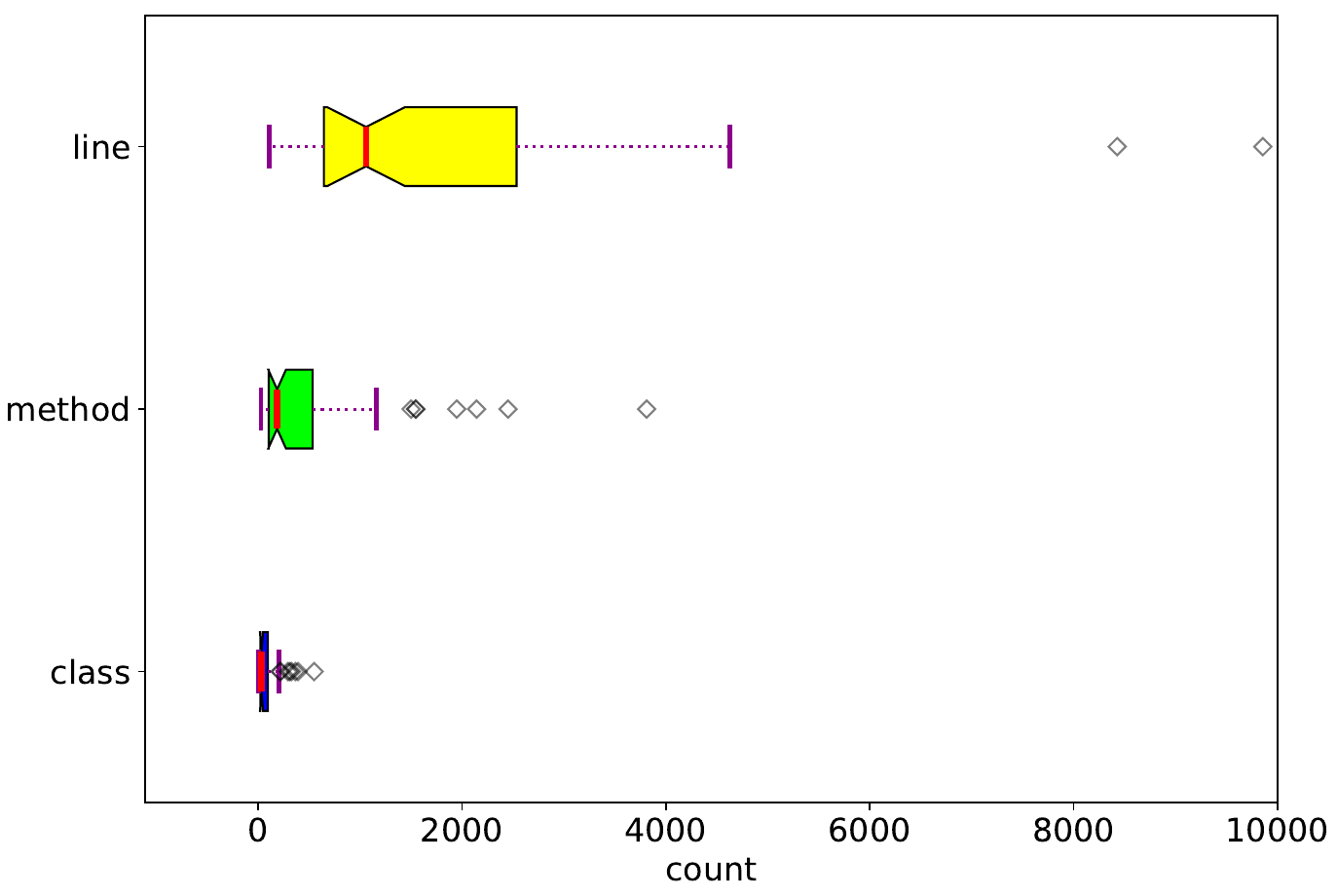}
\caption{Statistics of the benchmark apps}
\label{fig:stats}
\end{figure}

We evaluate \ourapproach based on AndroTest benchmark~\cite{choudhary2015automated}. It consists of 66 real-world apps, 
including popular apps like WordPress and Wikipedia. Five apps were excluded due to the failures, such as crashes and incompatibility issues, during the experiments with either our approach, \ape, or \deepgui. 
This benchmark was used for comparing various Andriod testing tools in~\cite{choudhary2015automated}, which includes Dynodroid (a random fuzzing tool)~\cite{machiry2013dynodroid}, GUIRipper (a model-based depth first search tool)~\cite{guiripper-ase12}, and EvoDroid (an evolutionary algorithm-based fuzzing tool)~\cite{mahmood2014evodroid} . 
It is also used in the experiments of Humanoid~\cite{li2019humanoid} and \deepgui~\cite{yazdanibanafshedaragh2021deep}.
\autoref{fig:stats} shows the statistics of the benchmark apps. It shows that the sizes of the apps range from small (100 LOCs) to large (22 kLOCs). Median size is 1060 LOCs. Number of methods range from 449 to 5329. Number of classes range from 26 to 187. {\redcolor Expanding the dataset is non-trivial because our current benchmarking apps were custom instrumented and built from source to enable precise measurement of class, method, and line coverage. This instrumentation requires access to the source code and modifications to the build process, making it challenging to include a broader set of apps, particularly closed-source or third-party applications that lack this level of access and control. While this instrumentation was necessary for measurement purposes, our approach does not require source code to test real-world apps.}

\subsection{Comparison with State-of-the-Art} 
For RQ1, we compare \ourapproach with two state-of-the-art Android fuzzing approaches --- \ape~\cite{gu2019practical} and \deepgui~\cite{yazdanibanafshedaragh2021deep}. We ran \deepgui, \ape, and \ourapproach on the \numapps apps for 60 minutes each. To ensure reliability, we repeated this entire testing process five times. We then compare the effectiveness of the approaches by analyzing their average coverage results.

There are other approaches, namely Monkey~\cite{monkey}, Stoat~\cite{su2017guided}, and Sapienz~\cite{mao2016sapienz}, which are widely used as benchmarks in the literature. We do not compare \ourapproach with them because \ape and \deepgui have already compared theirs with those approaches and outperformed them in their experiments.
In addition, Choudhary et al.'s empirical study~\cite{choudhary2015automated} reported that Monkey~\cite{monkey} achieved the best code coverage among other fuzzing tools like Dynodroid~\cite{machiry2013dynodroid}, GUIRipper~\cite{guiripper-ase12}, ACTEve~\cite{anand2012automated}, SwiftHand~\cite{choi2013guided}, EvoDroid~\cite{mahmood2014evodroid}, etc. 
We note that \deepgui requires data collection (collection of GUI images) and model training. It means that the tool may not work well when testing the app that has a GUI model different from the training data.

One state-of-the-art approach most closely related to our approach is GPTDroid~\cite{liu2024make}. However, the original implementation referenced in their paper was no longer available in the specified repository at the time of our evaluation. Our attempts to contact the authors for access to their implementation were unsuccessful. While we located an implementation sharing the same name on GitHub, we cannot verify if this represents the original work described in their paper. Therefore, we opted to provide a comparison of our approach and GPTDroid in related work (Section~\ref{sec:related}).

\subsection{Metric}
\emph{Coverage}: We use the standard coverage criteria --- class coverage, method coverage, and line coverage --- to assess the capability of the fuzzers. We shall also use the time vs the coverage as a metric for measuring the efficiency of the tools.

\subsection{Setup}
\label{sec:setup}
\ourapproach runs both on emulators and real devices. It requires neither instrumentation nor modification to the app under test, Android device, or Android operating system. It only requires the device to be connected to the computer running \ourapproach tool via Android Debug Bridge (ADB). Apart from Android SDK tools such as ADB and Android Asset Packaging Tool (AAPT), \ourapproach does not have other dependencies.

We ran \ourapproach on an Android Google Pixel 2 emulator with 4GB of RAM running Android 9 (API level 28) to ensure compatibility with \ape in our experiments. However, \ourapproach was tested on Android versions 4.1 - 12 and it operates seamlessly as a black-box testing tool out-of-the-box. We ran \deepgui on Android 2.3.3 (API level 10).
We monitored the class, method, and line coverage of the app under test every 60 seconds using Emma tool~\cite{rubtsov2006emma}.

All emulators were run on a Ubuntu Desktop Intel Core i9 64GB RAM @ 3.2 GHz, 64GB. We ran all tools with their default configurations.

\subsection{Results}
\label{sec:results}

This section discusses the experiment results addressing the research questions. 

\subsubsection{RQ1: Code Coverage}
\label{sec:rq1}
\begin{figure*}[h]
    \centering
    \includegraphics[width=0.95\textwidth]{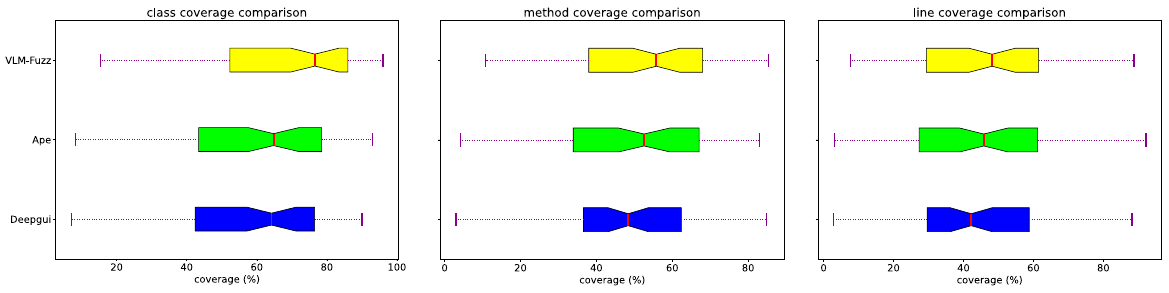}
    \centering
    \begin{tabular}{lccccc}
    \textbf{Coverage} & \textbf{\deepgui} & \textbf{\ape} & \textbf{\ourapproach} 
        & \textbf{\ourapproach}  & \textbf{\ourapproach}  \\ 
        
    & (\%) & (\%) & (\%) & vs \textbf{\deepgui} & vs \textbf{\ape} \\ \hline
    Class &  \deepclasscov &  \apeclasscov &  \classcov & p=0 & p=0   \\
    Method &  \deepmethodcov &  \apemethodcov &  \methodcov & p=0.002 & p=0.004 \\
    Line &  \deeplinecov &  \apelinecov &  \linecov & p=0.072 & p=0.037  \\
   
    \hline
    \end{tabular}
    \caption{Comparison of Class, Method, and Line Coverage, averaged across 5 runs (RQ1)}
    \label{fig:rq1a}
\end{figure*}

Figure~\ref{fig:rq1a} shows the boxplots and the means of covered classes, covered methods, and covered lines of \ourapproach and the baselines. In general, \ourapproach outperforms against the baselines across the three coverage criteria. On average, \ourapproach achieves class coverage of \classcov\%, method coverage of \methodcov\%, and line coverage of \linecov\%, across the \numapps apps. 
\ourapproach outperforms the baselines in terms of coverage. On average, compared to \deepgui and \ape, it achieved  \dclasscov and \aclasscov relative improvements in terms of class coverage, \dmethodcov and \amethodcov relative improvements in terms of method coverage, and \dlinecov  and \alinecov  relative improvements in terms of line coverage, respectively.~\autoref{tab:rq1b} shows the number of apps for which a given approach achieves a better coverage. It shows that \ourapproach performed better than both \deepgui and \ape on a significant number of apps (more than double) in terms of class and method coverage. It performed better than \deepgui and \ape on 42 and 35 out of 59 apps, respectively, in terms of line coverage.

We also computed the statistics (p-values) of paired t-tests comparing \ourapproach vs \deepgui and \ourapproach vs \ape. We used t-tests instead of non-parametric tests like Wilcoxon signed rank test, because t-test has higher statistical power, but it requires data distribution to be normal. We first computed the p-values of the Shapiro-Wilk test with respect to the difference in coverage results between \ourapproach and \deepgui, and between \ourapproach and \ape. All the p-values are 0.8 and above. If the p-value of the Shapiro-Wilk Test is greater than 0.05 (chosen alpha value), we can assume that the data is normally distributed. The last two columns in ~\autoref{fig:rq1a} show the statistical results. As the p-values are smaller than 0.05,  \ourapproach is statistically better than both \deepgui and \ape in terms of class and method coverage, and \ourapproach is also statistically better than \ape, in terms of line coverage.

\begin{table}[tbh]
\caption{Comparison based on \#apps for which the approach achieves better coverage (RQ1)}
\centering
  \begin{tabular}{lcccc}
      \textbf{Coverage}  & \textbf{\ourapproach}  vs \textbf{\deepgui} & \textbf{\ourapproach}  vs \textbf{\ape}  \\ \hline
     Class & 53 vs 6 & 49 vs 10 \\
     Method & 48 vs 11 & 40 vs 19 \\
     Line & 43 vs 16 & 35 vs 23 \\
    
    \hline
    \end{tabular}
\label{tab:rq1b}
\end{table}

\begin{figure}[tbh]
\centering
\includegraphics[width=0.47\textwidth]{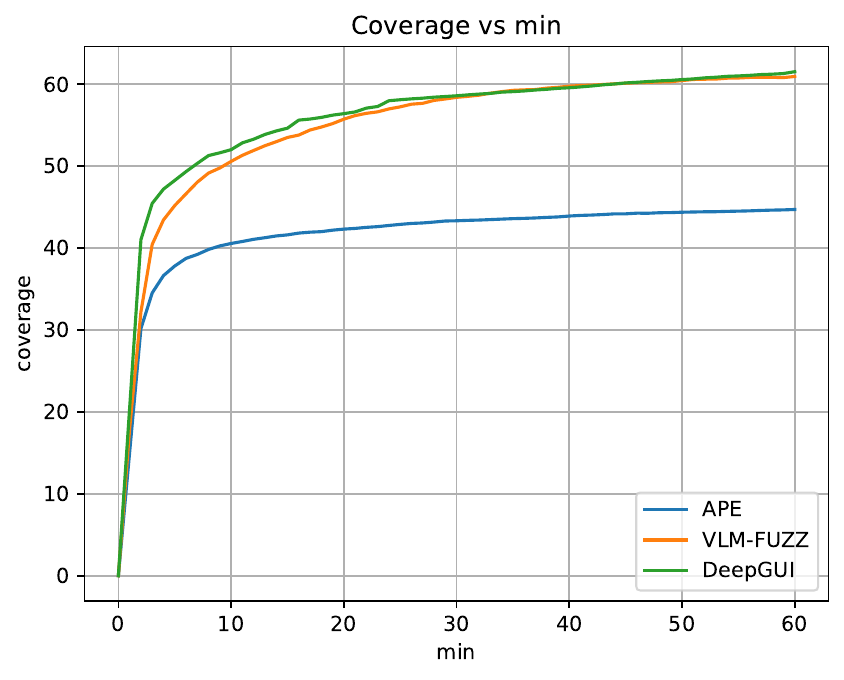}
\caption{Progressive average line coverage across all apps.}
\label{fig:coverage}
\end{figure}

Our experiments show that \ourapproach not only achieved a higher coverage on average but also reached to a high coverage point efficiently. As shown in Figure~\ref{fig:coverage}, which tracks the average coverage across all tested apps, \ourapproach reached 50\% code coverage in under 10 minutes; its coverage continues to improve over time and is yet to reach the peak. This can be contributed to our approach's clever time budget allocation to the GUI components according to their complexity. 

\begin{keytakeaway}[]{Key takeaway}
\verbatim
\ourapproach achieved 46.5\%, 53.2\%, 68.5\% line, method, and class coverages respectively on the benchmark apps, statistically outperforming the state-of-the-art approaches. 
\end{keytakeaway}

\subsubsection{RQ2: Ablation Study} 
Table~\ref{tab:rq2a} shows the number of apps that \ourapproach did not utilize its VLM module. From the total dataset of \numapps apps, \ourapproach was able to test \novision apps without using its VLM capabilities. It also shows that, on 9 out of \novision apps, non-VLM-assisted tests achieved better line coverage compared to the baseline tool, \ape. Note that in this ablation study, we compared with only \ape and based on line coverage. This is because the number of apps for which \ourapproach or \ape performed better in terms of line coverage is closer (35 vs 23) than the numbers from any other comparisons, as shown in~\autoref{tab:rq1b}. For example, in comparison with \deepgui, \ourapproach performed better for 43 out of 59 apps in terms of line coverage.

To understand the impact of the VLM component, we randomly selected \ablation apps for the ablation study from the remaining \usedvision apps that required VLM.
Table~\ref{tab:rq2b} shows the performance comparison of the non-VLM-assisted variant of \ourapproach against the full approach and the baseline, \ape.  
The results on these \ablation apps show that vision-based test generation is essential for reasoning with complex UI components to generate valid inputs. \ourapproach has a relative improvement of \novil line coverage compared to non-VLM-assisted variant of \ourapproach.

\noindent
{\bf \em Cost Efficiency and Scalability}: We computed the cost of using GPT-4o as our VLM module. Averaging across the \numapps apps we experimented with, it costs approximately US\$0.25 per app for one hour of fuzzing, which is reasonable for single-app tests. This is because \ourapproach is efficient in invoking VLM only when it is necessary, as explained in Section~\ref{sec:uianalyzer}. {\redcolor However, we recognize that this cost may increase significantly when scaling to large app sets or production environments. Moreover, depending on a proprietary model such as GPT-4o may affect reproducibility and long-term applicability because of LLM drift. In future work, we plan to explore cost optimization strategies and consider open-source alternatives to address these scalability concerns while maintaining consistent performance.}

\begin{keytakeaway}[]{Key takeaway}
\verbatim
By selectively applying a vision language model to complex GUIs while using heuristics and depth-first search for simpler interfaces, \ourapproach strikes a good balance between code coverage and cost, making it a cost-efficient solution for automated app testing.
\end{keytakeaway}

\begin{table}[h]
\caption{\#apps where \ourapproach does not utilize VLM}
\centering
  \begin{tabular}{cccc}
     No VLM & Won w/o VLM \\
    \hline
    13 & 9 \\
    \hline
    \end{tabular}
\label{tab:rq2a}
\end{table}

\begin{table}[h]
\caption{\ape vs With VLM (full approach) vs Without VLM, based on average line coverage (\%) (RQ2)}
\centering
  \begin{tabular}{lc|cc}
    Selected App & \ape & With VLM & Without VLM \\
    \hline
    \texttt{fsck} & 3.2 & 21 & 7.2 \\
    \texttt{yahtzee} & 10.6 & 53.8 & 45.8 \\
    \texttt{weightchart} & 24.2 & 59.8 & 57.2  \\
    \texttt{myLock} & 30.2 & 35.8 & 33.8 \\
    \texttt{fercanet} & 30.6 & 35.2 & 35.2 \\
    \hline
    \end{tabular}
\label{tab:rq2b}
\end{table}

\subsubsection{RQ3. Bug Detection Capability on Real-world Apps}
To evaluate the effectiveness of \ourapproach in detecting crash bugs, we conducted extensive testing on real-world apps. We selected \realworldapps official apps from the Google Play Store that were updated in 2024.
The selected apps represent a diverse user base, with download counts ranging from 500 to 1 billion (Table~\ref{tab:rq3b}).
Following the experimental setup detailed in Section~\ref{sec:setup}, we executed \ourapproach on this test suite. Our analysis focused specifically on fatal errors that caused app crashes, excluding errors that allowed processes to continue execution without crash. We prioritized fatal crashes because they strongly indicate the presence of software bugs and represent critical issues that directly impact user experience.
Not all runtime errors indicate software bugs, particularly those triggered by security mechanisms or invalid component access. For instance, apps implementing tamper-proofing controls typically throw \texttt{App Sealing} errors when executed in an emulator environment - a legitimate security response rather than a bug. Through manual inspection, we filtered out such security-related errors from our bug analysis.
We also observed that apps may throw runtime exceptions when a private activity component is invoked without the appropriate input data (such as private Intent parameters). While fuzzing tools can trigger these errors through privileged emulators and debugger mode, they would not occur during legitimate user interactions, and thus were not classified as software bugs. These considerations are different from existing approaches such as \cite{su2017guided, gu2019practical}. Although these issues were not classified as bugs, we note that implementing proper exception handling remains a best practice regardless of how components are accessed, as it improves overall app robustness.

\ourapproach induced crashes in \reportedbuggyapps apps. 
Two of the authors manually investigated the crash logs (generated via ADB logcat) of these apps, together with the screenshots of the apps generated during fuzzing. The authors confirmed that crashes in \realbuggyapps apps are caused by real bugs and that crashes in the other apps are due to the non-bug reasons stated above. In total, there are \numcrashes unique crashes in the buggy \realbuggyapps apps. We have sent the crash logs of those buggy apps to the respective developers. The logs are currently under investigation by the developers, with none of them rejected yet.

Table~\ref{tab:rq3a} shows the distribution of most common exception types we observed in the buggy apps. In total, we observed \exceptiontypes different types of exceptions that cause crashes. 
\texttt{NullPointerException} is the most common type of exceptions, which aligns to previous case studies~\cite{gu2019practical, su2017guided, mao2016sapienz}, followed by \texttt{RuntimeException} and \texttt{NullReferenceException}. 
Table~\ref{tab:rq3b} lists the name, version, and number of downloads of the buggy apps, the number of unique crashes and the crash types. 
We analyzed the details of these bugs. We observed that 9 of them involve multiple text inputs or compound operations to cause the crashes. Some crashes are only caused by particular sequences of operations. These findings indicate the ability of \ourapproach to explore the state space effectively and the effectiveness of our approach in bug detection.

\noindent
\textbf{\emph{Cost:}} We computed the cost of using GPT-4o by our VLM module for this experiment with real-world apps. Averaging across the \realworldapps official apps we experimented with, it costs approximately US\$0.05 per app.  

\begin{keytakeaway}[]{Key takeaway}
\verbatim
\ourapproach uncovered several unique crashes corresponding to \exceptiontypes different types of exceptions in \realbuggyapps real-world apps that are released recently in 2024. This demonstrates the effectiveness and applicability of our approach over latest Android versions.
\end{keytakeaway}

\begin{table}[h]
\caption{Distribution of unique fatal exceptions}
\centering
  \begin{tabular}{cc}
    Exception Type & Number \\
    \hline
    \texttt{NullPointerException (NPE)} & 46 \\
    \texttt{RuntimeException (RE)} & 27 \\
    \texttt{NullReferenceException (NRE)} & 16 \\
    \texttt{IllegalArgumentException (IE)} & 8 \\
    \texttt{IllegalStateException (ISE)} & 8 \\
    \texttt{InvocationTargetException (ITE)} & 7 \\
    \texttt{InflateException (IFE)} & 7 \\
    \texttt{ClassNotFoundException (CNFE)} & 6 \\
    \texttt{uncaughtException (UE)} & 6 \\
    \texttt{ArgumentNullException (ANE)} & 5 \\
    \texttt{Other exceptions (Others)} & 72 \\
    \hline
    \end{tabular}
\label{tab:rq3a}
\end{table}

\begin{table}[h]
\caption{Crashes detected in official apps (RQ3).  }
\centering
  \begin{tabular}{lcccc}
    App & Version & \#Download & \#Crash & Excep. Types \\
    \hline
    \texttt{facebook.lite} & 407.0.0.12.116 & 1B & 3 & ISE, Others \\
    \texttt{walmart} & 24.38.1 & 100M & 5 & IE, ISE, Others \\
    \texttt{myfitnesspal} & 24.37.0	& 100M & 25 & NPE, ITE, UE, \\
        & & & & IFE, Others \\
    \texttt{nike.omega} & 24.45.1 & 50M & 24 & NPE, RE, ITE, \\ 
        & & & & CNFE, ISE, Others \\
    \texttt{doordash} & 15.184.13 & 50M & 20 & IFE, ITE, RE, \\
        & & & & IE, Others \\
    \texttt{parallelspace} & 5.1.7 & 10M & 5 & CNFE, RE, \\ 
        & & & & NPE, Others   \\
    \texttt{ktshow} & 07.00.09 & 10M & 13 & RE, NPE, Others \\
    \texttt{hostelworld} & 9.62.0 & 5M & 4 & IFE, ITE, RE, IE  \\
    \texttt{avatrade} & 143.4 & 1M & 1 & NPE \\ 
    \texttt{lifebear} & 4.6.7 & 1M & 39 & NRE, RE, NPE,     \\
        & & & & CNFE, ANE, Others  \\        
    \texttt{myschoolbucks} & 13.5.0.122703 & 1M & 1 & NRE \\
    \texttt{baqeyat} & 3.3 & 100K  & 14 & RE, NPE, Others \\
    \texttt{obrien.dave} & 4.9.1 & 50K & 4 & ITE, IE, RE, Others \\
    \texttt{ggmgastro} & 5.59.4 & 10K & 3 & NRE, Others \\
    \texttt{adaptica} & 2.32 & 10K & 1 & NRE \\
    \texttt{passmobile} & 3.7 & 5K & 1 & NPE  \\
    \texttt{eparking.pam} & 2.0.5 & 1K & 1 & CNFE \\
    \texttt{jeulin.pcr} & 1.3.2 & 1K & 1 & NRE  \\ 
    \texttt{addsecure} & 5.2.1 & 1K & 1 & NRE  \\
    \texttt{forza} & 8 & 1K & 12 & RE, NPE, ISE, Others \\
    \texttt{windeditlite} & 3.0.0 & 1K & 5 & NRE, ANE, RE, Others   \\
    \texttt{jstay} & 1.27 & 1K & 1 & NRE   \\
    \texttt{shopgate} & 5.59.1 & 1K & 9 & RE, NPE,   \\
        & & & & UE, Others \\
    \texttt{StudioScientifique} & 1.1.3 & 500 & 15 & NRE  \\
    \hline
    \end{tabular}
\label{tab:rq3b}
\end{table}

\subsection{Case Studies}
\label{sec:casestudy}
To analyze \ourapproach's coverage patterns, we conducted detailed investigations of \sampleapps representative apps. Our selection considers three scenarios: apps where \ourapproach outperformed the best baseline, apps where it underperformed compared to the baseline, and apps where it operated without VLM assistance. This diverse sampling enables comprehensive understanding of the tool's effectiveness across different scenarios.

\texttt{bookcatalogue}: This is the running example app used in motivating and explaining our approach. For this app, \ourapproach achieved 5\% better line coverage than \ape. This app required VLM since it has challenging UIs that require reasoning about the semantic meanings of certain objects. When the app is first opened, the main activity component shows an empty bookshelf with no obvious interactive elements, except for a menu option at the bottom to add books. In another component called \texttt{BookEdit}, there is a group of input fields such as rating, notes, and dates. \ape seems to have generated random values for those fields whereas \ourapproach activated the VLM to generate inputs that have semantic meanings for each of the input fields.

\texttt{timer}: For this app, \ourapproach achieved 8\% better line coverage than \ape. This app required VLM. The app has timer controls like setting minutes and pausing or canceling the timer. Similar to the above, \ourapproach activated the VLM to generate inputs that have semantic meanings for those inputs.

\texttt{weightchart}: For this app, \ourapproach achieved 25\% better line coverage than \ape. This app required VLM. \ourapproach handled input fields, such as height and weight in \texttt{ChartActivity} and \texttt{EntryActivity} components, with more accuracy than \ape. In addition, \ape also missed exploring certain states in those activities whereas \ourapproach managed to explore them efficiently, thereby, giving \ourapproach an advantage in code coverage.

\texttt{blinkenlights}: For this app, \ourapproach did not invoke Vision LM since its UI does not have complex graphical elements. There is only one major activity component to test. Its UI is not complex but has a setting feature that contains multiple interactive elements like checkboxes and theme selection. \ourapproach achieved a significantly higher line coverage than the baseline ($>$25\%), which shows the capability of our heuristic-based state space exploration algorithm to reach deeper codes within the app's component.  

\texttt{divideandconquer}: For this app, \ape achieved $>$6\% line coverage than \ourapproach. This was mainly because our tool does not support the ``drag'' user action required to test a component called \texttt{DivideAndConquerActivity}. The app requires the user to play a game by dragging the lines. \ape managed to generate the drag action each time, thereby covering more of the app's functionality.  
The UI actions currently supported in our implementation are stated in Section~\ref{sec:implementation}. 

\texttt{multismssender}: For this app, \ape achieved $>$15\% line coverage than \ourapproach. This was due to the similar reason mentioned above.

Overall, we observed that \ourapproach performs better when the app requires complex interactions with the user, the system, and/or other apps and whether it requires reasoning about a group of UI elements to determine the required inputs.
We also observed that while VLMs excel at identifying visually apparent interface elements, they lack in generating non-visible interaction patterns, such as long-press gestures, scroll behaviors, and menu activations. Furthermore, the computational complexity and associated costs of iterative VLM queries following each interaction event, similar to the approach in \cite{zhang2023appagent,liu2024vision}, render this approach suboptimal for comprehensive test coverage. The experimental results suggest that an optimal testing methodology should integrate VLM capabilities with conventional heuristic-based testing algorithms, thereby establishing a more robust framework for UI testing.

\begin{keytakeaway}[]{Key takeaway}
\verbatim
{\redcolor 
While our approach builds on existing exploration strategies, the integration of VLMs introduces a novel way to enhance UI reasoning and input generation. The improvements in code coverage are statistically significant and demonstrate the potential of VLM-assisted fuzzing. Furthermore, our method also demonstrates bug detection capabilities, effectively identifying different issues within real-world apps. Future optimizations aim to further enhance efficiency while balancing complexity and performance gains.}
\end{keytakeaway}

\subsection{Threat to Validity}
\label{sec:threats}
We adopted several strategies to mitigate the threats to the validity of our experimental results.

{\em Internal validity} threats concern external factors affecting the independent variable. To mitigate the risk of bias in selecting the case studies, we chose apps from an existing benchmark already used in previous studies. Additionally, the VLM used was proprietary, which might have evolved during the experiment (LLM drift), i.e., the model in the first part of the experiment might differ from the model in the last part of the experiment. We mitigated this threat by running our experiment for a short period of time, thus the model evolution should have been limited or none.

{\em Construct validity}  threats concern the relationship between theory and observation. We controlled these threats by using Emma tool~\cite{rubtsov2006emma} to measure code coverage, which is a standard coverage tool widely used in literature such as~\cite{choudhary2015automated,su2017guided,romdhana2022deep}.

{\em Conclusion validity} threats concern the relationship between treatment and outcome. We used established statistical tests, Shapiro-Wilk test and paired t-test, to draw conclusions only when statistical significance was detected.

Eventually {\em external validity} concerns the generalization of the findings. Our results might not generalize to other case studies or mobile architecture (e.g., iOS). We controlled this threat by considering a large number of apps, but only further experiments with more apps and devices can confirm our results.

\section{Related Work}
\label{sec:related}
One state-of-the-art approach that is most closely related to our approach is GPTDroid~\cite{liu2024make}. Since GPTDroid tool is not available to us, in the following, we base our comparative discussion primarily on the methodological description provided in their paper, focusing on the conceptual differences between our approaches. 

State tracking is managed by combining three key pieces of information: a summary of the last k=5 steps leading up to the current state,  the number of times a widget is accessed and the name of the current component. Using this state information, the LLM determines what action to take next. However, this simplified approach to state tracking has limitations, particularly when handling complex UI interactions. The problem becomes evident in scenarios where every action changes the main UI. For instance, when the test starts on Activity A (State A1), opens a popup, returns to a modified version of Activity A (State A2), opens another popup, and then returns to yet another version of Activity A (State A3) that is equivalent to the first state (State A1), the system may struggle to differentiate between these states. Without more sophisticated state management, this can result in the approach getting trapped in an infinite loop as it fails to recognize the subtle differences between the various versions of the same UI. 

XML layout hierarchy attributes alone have significant limitations when interpreting user interface elements for input generation - which GPTDroid is based on. One major challenge is missing textual data – our analysis of 3,831 unique non-layout widget instances across the entire test apps revealed that 34.9\% of the instances miss {\em text} attributes, 40.6\% miss {\em resource-id}, and 24.5\% are missing both making it challenging developing a heuristic based on these attributes. Table~\ref{tab:widget-insignt} summarizes the findings.
Widget interactivity introduces additional complexity: elements that might be mistakenly considered non-interactive because {\em all} their boolean attributes (such as {\em "clickable"} or {\em "long-clickable"}) are set to {\texttt false} can still be interactive as they may inherit interactive properties from their parent containers. Additionally, the use of identical {\em resource-id} attributes across multiple widgets in the same UI creates identification ambiguity. This limitation becomes particularly evident in scenarios requiring visual reasoning, such as with password confirmation fields. In these cases, the approach needs to understand that it must enter identical text in both the original and confirmation fields - a relationship that may not be apparent from the XML layout attributes alone, especially if the widget attributes lack descriptive details. \ourapproach, however, addresses these limitation by combining heuristic-based depth-first state exploration with VLM-assisted event generation.

\begin{table}
\centering
\caption{Widget attribute insight across the test apps}
\begin{tabular}{l|c}
\textbf{Attribute} & \textbf{\%} \\
\hline
text \ding{55}, resource-id \ding{51} & 18.8\% \\
text \ding{55}, resource-id \ding{55} & 16.1\% \\
text \ding{51}, resource-id \ding{55} & 24.5\% \\
text \ding{51}, resource-id \ding{51} & 40.6\% \\
\hline
text \ding{55} & 34.9\% \\
resource-id \ding{55} & 40.6\% \\

\hline
\end{tabular}
\label{tab:widget-insignt}
\end{table}

Different from \ourapproach, GPTDroid does not generate system Intents, hence, fails to simulate how an app might respond to system-wide broadcasts, potentially missing crucial functionality that relies on these events. Moreover, not considering intent-filters means it cannot adequately test various app launch scenarios or deep-linking capabilities. Finally, its limitation in simulating menu tap actions that could potentially trigger pop-up menus could result in incomplete UI testing. 

{\redcolor Trident~\cite{liu2024vision} 
uses multimodal large language models to detect non-functional bugs by reasoning with the visual and functional logic of GUI pages. Different from ours, Trident purely relies on LLMs for both test generation and state space exploration. Our approach integrates static analysis, DFS, and a vision-language model for systematic state exploration. Since Trident has not been peer-reviewed (arXiv paper) yet, we did not compare with Trident in our experiments. But from our case studies (Section~\ref{sec:casestudy}), we have observed that an optimal testing methodology should integrate VLM capabilities with conventional heuristic-based testing algorithms, for a more robust testing framework.} 

There are many other Android GUI fuzzing approaches that focus on improving code coverage.
Choudhary et al. (Androtest)~\cite{choudhary2015automated} conducted a comparison of \emph{automated} test input generation tools for Android. 
The compared approaches include Monkey~\cite{monkey}, Dynodroid~\cite{machiry2013dynodroid}, Droidfuzzer~\cite{ye2013droidfuzzer}, Intent fuzzer~\cite{sasnauskas2014intent}, GUIRipper~\cite{guiripper-ase12}, Orbit~\cite{yang2013grey}, SwiftHand~\cite{choi2013guided}, Puma~\cite{hao2014puma}, A3E-Targeted~\cite{azim2013targeted}, EvoDroid~\cite{mahmood2014evodroid}, and ACTEve~\cite{anand2012automated}. In terms of the state space exploration strategy, these approaches employ either random exploration or model-based exploration.  In model-based approaches, state space exploration is guided by a model of the app. For example, SwiftHand~\cite{choi2013guided} uses finite state machine model of the app and A3E-Targeted~\cite{azim2013targeted} uses Static Activity Transition Graph of the app to guide the exploration of activity components and transitions. They may also employ a search strategy such as depth first search~\cite{guiripper-ase12, yang2013grey}, evolutionary search~\cite{mahmood2014evodroid}, or symbolic execution~\cite{anand2012automated}. 

To further improve the scalability and test coverage, some approaches such as Sapienz~\cite{mao2016sapienz} combines random fuzzing and search-based test generation.
DroidBot~\cite{li2017droidbot} proposes a lightweight UI-guided test input generator that can generate UI-guided test inputs based on a state transition model generated dynamically. 
Stoat~\cite{su2017guided} uses dynamic analysis to construct a stochastic finite state machine based on only UI events, for which a probability is assigned and updated based on their execution frequency. Next, it generates test sequences with UI events and randomly injected system-level events.  

To address the problem of incomplete model of the app when built statically, APE~\cite{gu2019practical} applies both static and dynamic analyses. It first builds a static GUI model (finite state machine) and updates the model during testing. It generates test cases based on the model to explore various states/transitions/sequences. 

In general, random exploration strategy often fails to create reasonable testing paths tailored to the app’s characteristics, leading to low test coverage. While model-based exploration is systematic and can enhance test coverage, its code coverage can be further improved by reasoning with the semantic information of the app’s GUI and its widget items.

Since machine learning techniques are capable of learning from data, they have been incorporated into model-based exploration, to further improve the test coverage. For example, DeepGUI~\cite{yazdanibanafshedaragh2021deep} complements fuzzing with a more intelligent form of GUI input generation. Given screenshots of apps, Deep GUI first employs deep learning to construct a model of valid GUI interactions. It then uses this model to generate effective inputs, given screenshots from the app under test.

However, the above-mentioned approaches may still struggle in dealing with complex and dynamic layout of activity screens, reasoning about the context of certain widget items, and  getting stuck in the same activity screen or keep generating similar inputs that cause the app to crash or exit due to the lack of coordination between current input and previous inputs. These challenges were demonstrated in our experiments with existing approaches. In addition, these approaches do not consider the complexity of the components for allocating time budget for testing. In our experiments, we demonstrated that clever budgeting of testing time to complex components (those with several interactive UI elements) result in better code coverage.

\section{Conclusion}
\label{sec:conclusion}
In this paper we present \ourapproach, a novel approach to Android app UI fuzzing that uniquely combines Vision Language Models with traditional heuristic-based exploration strategies. Our work addresses critical limitations in existing Android testing approaches, particularly the challenge of achieving high code coverage when dealing with complex UI interactions. By leveraging the visual reasoning capabilities of VLMs alongside a recursive depth-first search exploration strategy, \ourapproach demonstrates significant improvements over current state-of-the-art solutions.
The experimental results across the test apps show that \ourapproach achieves better performance metrics, showing notable improvements over the best baseline approach. This performance enhancement validates our hypothesis that incorporating visual reasoning capabilities through VLMs can significantly improve the effectiveness of automated Android UI testing.
Our ablation studies further confirm the value of integrating VLMs into the testing process, demonstrating that their ability to reason about complex GUI objects contributes significantly to improved test coverage. The results of \ourapproach suggests that the combination of traditional search-based strategies with advanced AI models represents a promising direction for future research in automated software testing.
By making \ourapproach publicly available, we provide a practical tool that can be immediately used by the software engineering community. Future work could explore extending this approach to other mobile platforms or web apps and pretraining/fine-tuning a specialized vision model, or investigating the integration of more advanced VLMs as they become available.
The results of this study demonstrate that combining classical software testing methodologies with modern AI capabilities can lead to more effective and efficient testing solutions.

\bibliographystyle{plainnat}
\bibliography{reference}

\appendix
\section*{Appendix}
\section{UI interaction algorithms}
\label{sec:appendix}

\subsection{performVisionActions()}
Algorithm~\ref{alg:perform-action-vision} outlines the \texttt{performVisionActions()} function. The function begins by capturing a screenshot of the current UI. It then labels the interactive widgets based on the dynamically retrieved UI hierarchy. Next, it queries the VLM to generate action sequences. The function iterates through these actions, executing each one on its corresponding widget. When an action triggers a UI change, the function instantiates a new \texttt{UI\_Analyzer()} to process the components in the updated interface.
 
\begin{sansserifalgoritm}
\caption{Perform Vision Generated Actions }
\begin{algorithmic}[1]
\small
\Statex
\Statex \textcolor{bluetext}{\textbf{Input:}} Inferred interactive widgets list ($view\_items\_in\_component$)
\Function{performVisionActions}{$widgetsList$}

    \State $screenshot \gets \text{takeScreenshot}()$

    \State $labelledScreenshot \gets \text{labelScreenshot}(screenshot)$

    \State $actionSequence, summary = \text{getVisionActions}(labelledScreenshot)$

    \State $actionsList \gets \text{getActions}(actionSequence)$
    
    \For{$action \in actionsList$}
        \If{$\text{getCurrentComponent}() != currentVisibleUI()$}
            \If {$\text{\textbf{\textcolor{bluetext}{not }}} \text{replay}()$}
                \State $\text{\textcolor{OliveGreen}{// we couldn't backtrack}}$                
                \State \Return
            \EndIf
        \EndIf

        \State $label \gets \text{getLabel}(action)$
        \State $widget \gets \text{getWidgetWithLabel}(label)$
        \If{$\text{"input"} \in action$}
            \State $text \gets \text{getText}(action)$
            \State $\text{sendText}(widget, text)$
        \EndIf

        \If{$\text{"tap"} \in action$}    
            \State $\text{sendTapOrLongPress}(widget)$
        \EndIf

        \If{$\text{"scroll"} \in action$}    
            \State $scrollDirection \gets \text{getDirection}(action)$
            \State $\text{scroll}(scrollDirection)$
        \EndIf
        \If{$\text{actionCausedUiChange}()$}
            \State $currentComponent \gets \text{getCurrrentComponent}()$
            \State $\text{UI\_Analyzer}(currentComponent)$ 
        \EndIf

    \EndFor

    \State $\text{tap\_menu}()$ 

    \State $\text{rotateAndRestoreScreen}()$ 

    \If{$\text{actionCausedUiChange}()$}
        \State $currentComponent \gets \text{getCurrrentComponent}()$
        \State $\text{UI\_Analyzer}(currentComponent)$ 
    \EndIf

    \State \Return 
\EndFunction

\end{algorithmic}
\label{alg:perform-action-vision}
\end{sansserifalgoritm}

\subsection{performNonVisionActions()}

Algorithm~\ref{alg:perform-action-non-vision} outlines the \texttt{performNonVisionActions()} function. The function  traverses the list of widgets in the current UI. For each identified widget, it executes its associated action based on predefined mappings. If any action results in a UI state change, the function initiates a new \texttt{UI\_Analyzer()} instance to evaluate the components within the modified interface.

\begin{sansserifalgoritm}
\caption{Perform Non-Vision Generated Actions}
\begin{algorithmic}[1]
\small
\Statex
\Statex \textcolor{bluetext}{\textbf{Input:}} Inferred interactive widgets list ($widgetsList$)
\Function{performNonVisionActions}{$widgetsList$}

    \For{$widget \in widgetsList$}
        \If{$\text{getCurrentComponent}() != currentVisibleUI()$}
            \If {$\text{\textbf{\textcolor{bluetext}{not }}} \text{replay}()$}
                \State $\text{\textcolor{OliveGreen}{// we couldn't backtrack}}$
                \State \Return
            \EndIf
        \EndIf

        \If{$\text{widget.action} == ACTION\_TEXT$}
            \State $text \gets \text{getRandomOrLLMText}()$
            \State $\text{sendText}(widget, text)$
        \EndIf   

        \If{$\text{widget.action} == ACTION\_TAP$}
            \State $\text{sendTapOrLongPress}(widget)$
        \EndIf   

        \If{$\text{widget.action} == ACTION\_SCROLL$}
            \State $\text{scrollDown}(widget)$
        \EndIf 
        
        \If{$\text{actionCausedUiChange}()$}
            \State $currentComponent \gets \text{getCurrrentComponent}()$
            \State $\text{UI\_Analyzer}(currentComponent)$ 
        \EndIf

    \EndFor

    \State $\text{tap\_menu}()$ 

    \State $\text{rotateAndRestoreScreen}()$ 

    \If{$\text{actionCausedUiChange}()$}
        \State $currentComponent \gets \text{getCurrrentComponent}()$
        \State $\text{UI\_Analyzer}(currentComponent)$ 
    \EndIf

    \State \Return 
\EndFunction
\end{algorithmic}
\label{alg:perform-action-non-vision}
\end{sansserifalgoritm}

\end{document}